\documentclass[aps,prb,reprint,
superscriptaddress,%
twocolumn,showpacs,%
amsmath,amssymb]{revtex4-1}
\usepackage{graphicx}
\usepackage{dcolumn}
\usepackage{bm}
\usepackage{color}
\usepackage{graphicx}
\usepackage{upgreek}
\begin{document}
\title{Spin dynamics in a Curie-switch}
\author{A.~F.~Kravets}
\email{anatolii@kth.se}
\thanks{Correspondence author}
\affiliation{Institute of Magnetism, National Academy of Sciences of Ukraine, 36 b Vernadsky Blvd., 03142 Kyiv, Ukraine}
\affiliation{Nanostructure Physics, Royal Institute of Technology, 10691 Stockholm, Sweden}
\author{A.~I.~Tovstolytkin}
\affiliation{Institute of Magnetism, National Academy of Sciences of Ukraine, 36 b Vernadsky Blvd., 03142 Kyiv, Ukraine}
\author{Yu.~I.~Dzhezherya}
\affiliation{Institute of Magnetism, National Academy of Sciences of Ukraine, 36 b Vernadsky Blvd., 03142 Kyiv, Ukraine}
\author{D.~M.~Polishchuk}
\affiliation{Institute of Magnetism, National Academy of Sciences of Ukraine, 36 b Vernadsky Blvd., 03142 Kyiv, Ukraine}
\affiliation{Nanostructure Physics, Royal Institute of Technology, 10691 Stockholm, Sweden}
\author{I.~M.~Kozak}
\affiliation{Institute of Magnetism, National Academy of Sciences of Ukraine, 36 b Vernadsky Blvd., 03142 Kyiv, Ukraine}
\author{V.~Korenivski}
\affiliation{Nanostructure Physics, Royal Institute of Technology, 10691 Stockholm, Sweden}
\date{\today}
\begin{abstract}
Ferromagnetic resonance properties of F$_1$/f/F$_2$/AF multilayers, where weakly ferromagnetic spacer f is sandwiched between strongly ferromagnetic layers F$_1$ and F$_2$, with F$_1$ being magnetically soft and F$_2$ -- magnetically hard due to exchange pinning to antiferromagnetic layer AF, are investigated. Spacer-mediated exchange coupling is shown to strongly affect the resonance fields of both F$_1$ and F$_2$ layers. Our theoretical calculations as well as measurements show that the key magnetic parameters of the spacer, which govern the ferromagnetic resonance in F$_1$/f/F$_2$/AF, are the magnetic exchange length ($\Lambda$), effective saturation magnetization at $T=0$ $(m_0)$, and effective Curie temperature ($T_{\text{C}}^{\text{eff}}$). The values of these key parameters are deduced from the experimental data for multilayers with f = Ni$_x$Cu$_{100-x}$, for the key ranges in Ni--concentration ($x=54\div70$ at. \%) and spacer thickness ($d=3\div 6$ nm). The results obtained provide a deeper insight into thermally-controlled spin precession and switching in magnetic nanostructures, with potential applications in spin-based oscillators and memory devices.
\end{abstract}
\pacs{76.50.+g, 75.30.Et, 75.70.-i, 85.70.Kh}
\maketitle
\section{Introduction}

Spin valves, whose central functional part contains two ferromagnetic layers (F$_1$, F$_2$) separated by a nonmagnetic spacer, have been the foundation for a wide range of applications in nanoelectronics and spintronics.\cite{sv1,1,2} Recent studies have demonstrated that incorporation of diluted ferromagnetic layer (f) instead of the nonmagnetic spacer layer may expand the functionality of the spin valves, yielding nanostructures with thermally-controlled magnetic properties, of F$_1$/f/F$_2$ generic type.\cite{3,4,5,6} In such structures, the exchange coupling between strongly ferromagnetic outer layers F$_1$ and F$_2$ depends on whether the temperature is higher or lower than the effective Curie temperature of the spacer ($T_{\text{C}}^{\text{eff}}$). At low temperatures,  $T<T_{\text{C}}^{\text{eff}}$, the direct exchange interaction through the spacer in its ferromagnetic state favors the parallel orientation of the magnetic moments $M_1$ and $M_2$ of the outer layers, F$_1$ and F$_2$. At high temperatures, $T>T_{\text{C}}^{\text{eff}}$, $M_1$ and $M_2$ are exchange decoupled and their orientations can be changed independently by applying a suitable external magnetic field, $H$. Thus, a variation in temperature and/or field can produce switching between the parallel (P) and antiparallel (AP) mutual orientations of $M_1$ and $M_2$ in the system.\cite{7,8}

The key element in the F$_1$/f/F$_2$ sandwich described above (the so-called \textit{Curie-switch} or \textit{Curie-valve}) is the weakly ferromagnetic spacer, f, which should have a narrow ferromagnetic-to-paramagnetic transition and have the $T_{\text{C}}^{\text{eff}}$ value tunable in fabrication. Diluted ferromagnetic alloys, such as Ni-Cu, is the natural choice for the spacer material, since the Curie temperature of bulk \cite{9} as well as film \cite{10,11,12,13} samples of Ni$_x$Cu$_{100-x}$ alloys depends almost linearly on Ni concentration.

The experiments described in Refs.~\onlinecite{7,8} confirmed the concept of temperature-controlled P to AP switching in nanostructures F$_1$/f/F$_2$, in particular containing a Ni$_x$Cu$_{100-x}$ ($x=35\div72$ at.~\%) spacer enclosed by an exchange-pinned Co$_{90}$Fe$_{10}$ layer and a free Ni$_{80}$Fe$_{20}$ (Py) layer: Py/Ni$_x$Cu$_{100-x}$/Co$ _{90}$Fe$_{10}$/Mn$_{80}$Ir$_{20}$ (hereinafter -- F$_1$/f/F$_2$/AF, AF denoting antiferromagnetic Mn$_{80}$Ir$_{20}$). Since the earlier work primarily aimed at understanding the switching effect itself, little attention was paid to the effect of the spacer-mediated exchange on the ferromagnetic resonance in the structure.

This work investigates the magnetic resonance properties of the Curie-switch, experimentally and theoretically, aiming at understanding the mechanisms involved and obtaining the intrinsic physical parameters governing the spin dynamics in the system.
 
\section{Theory}

Consider an F$_1$/f/F$_2$/AF multilayer outlined above, where a weakly ferromagnetic spacer (f) is sandwiched between magnetically soft (F$_1$) and hard (F$_2$) layers, with F$_2$ exchange pinned by an antiferromagnetic layer (AF). The thicknesses of F$_1$, F$_2$ and f are $l_1$, $l_2$, and $d$, respectively.

Our calculations of the magnetic resonance fields will assume that F$_1$ and F$_2$ are single domain. For small layer thicknesses and strong exchange interaction, in the weak excitation limit typical of ferromagnetic resonance (FMR) experiments, this assumption is well justified.\cite{14} Spacer f with magnetization $m$ provides a relatively weak coupling between the outer ferromagnets, F$_1$ and F$_2$. The analysis aims to determine the effect of this interlayer exchange coupling, variable in strength as a function of temperature, on the FMR.
 
\begin{figure}[t,p]
\begin{center}
\includegraphics[width=6.5cm]{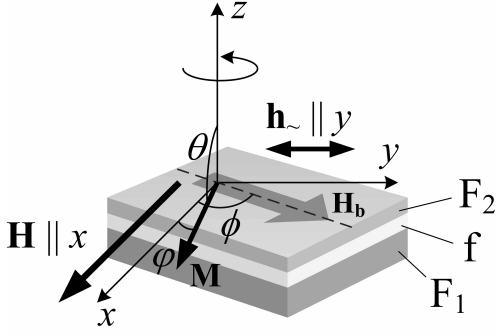}
\end{center}
\caption{(Color online)~Schematic of F$_1$/f/F$_2$/AF multilayer and chosen coordinate system in FMR measurements.}
\label{fig1}
\end{figure}

We use the classical Landau-Lifshitz approach \cite{15,16} to describe the F$_1$/f/F$_2$/AF multilayer and focus on the case where the quasistatic external field $H$ and the alternating field $h$ are in the film plane. Axes $Ox$ and $Oy$ are directed along $H$ and $h$, respectively (Fig.~\ref{fig1}). With $Oz$ perpendicular to the film plane, the magnetization vectors of F$_1$ and F$_2$ can be expressed as
\begin{equation}\label{eq:1}
\textbf{M}_i=M_i(\sin\theta_i\cos\varphi_i,\sin\theta_i\sin\varphi_i, \cos\theta_i),
\end{equation}
where $i=1,2$; $M_i$ is the saturation magnetization of $i$--th layer; $\theta_i$ and $\varphi_i$ are the polar and azimuthal angles, respectively.

The exchange bias between F$_2$ and AF can be modelled using an effective biasing field $H_{\text{b}}$ acting only on magnetization $M_2$.\cite{17} It will be shown below that for fully describing the FMR-effects of interest in this work, it is sufficient to consider only two cases, for which the external field $H$ is either parallel or antiparallel to $H_{\text{b}}$. Correspondingly, the value of $H_{\text{b}}$ can be either positive ($\phi=0$ in Fig.~\ref{fig1}) or negative ($\phi=\pi$) along the biasing axis. 

The magnetic energy of the $i$-th ferromagnetic layer in the above geometrical notations is
\begin{subequations}
\begin{eqnarray}\label{subeq:2a} 
W_i=Sl_i w_i \;,
\end{eqnarray}
\begin{multline}\label{subeq:2b}
w_i=2\pi M_i^2\cos^2\theta_i-M_iH_i\cos\varphi_i\sin\theta_i \\
-M_i h\sin\varphi_i\sin\theta_i \;,
\end{multline}
\end{subequations}
where $S$ is the area of the film surface, $w_i$ is the $i$-th layer energy density, $H_1=H$, and $H_2=H+H_{\text{b}}$. The first terms in Eq.~(\ref{subeq:2b}) originates from demagnetizing energy, while the second and third terms describe the energies of the interaction of the layers' magnetizations with the quasi-static $H$ and alternating $h$ external magnetic fields, respectively.

To simplify the equations, let us recall that in the case of a thin film, its high out-of-plane demagnetization fields prevent the magnetization vector from strongly deviating from the $xOy$ plane. In this case, $\theta_i$ can be represented as $\theta_i=\pi /2+\varepsilon_i$, where $\vert\varepsilon_i\vert \ll 1$.

In the small-signal approximation relevant for FMR, the magnetization vectors of F$_1$ and F$_2$ are nearly aligned with the $Ox$ axis (the easy axis, also the direction of external field $H$) and perform only weak oscillations near the ground state under the microwave excitation $h$. This means that $\vert\varphi_i\vert \ll 1$. The limits of validity of this approximation will be discussed in the experimental section below.

In the above small-signal thin-film approximation, the magnetic energy density of the system becomes
\begin{multline}\label{eq:3}
w_i=-M_iH_i+(4\pi M_i^2+M_iH_i)\varepsilon_i^2 /2 \\
+M_iH_i\varphi_i^2 /2-M_ih\varphi_i\;.
\end{multline}

F$_1$ and F$_2$ are exchange coupled through a weakly ferromagnetic spacer f. The case where the spacer is highly magnetically diluted and nominally (in the bulk) is paramagnetic was considered in Ref.~\onlinecite{8}, with the interlayer exchange mediated via induced proximity ferromagnetism. Here we consider the case where the spacer is diluted such that it is nominally ferromagnetic and can mediate direct exchange between the outer ferromagnetic layers, with the exchange coupling strength being a steep function of temperature near the Curie point of the spacer. 

We denote the temperature dependent saturation magnetization of the spacer as $m$. Assuming again that the magnetization is uniform in the $xOy$ plane, the spacer energy density can be written as
\begin{multline}\label{eq:4}
w=\dfrac{\alpha m^2}{2} \left[\left(\dfrac{\partial\theta}{\partial z}\right)^2 + \sin^2 \theta \left(\dfrac{\partial\varphi}{\partial z}\right)^2 \right] \\
-\dfrac{\textbf{m}\textbf{H}_m}{2}-Hm \cos\varphi\sin\theta-h m \sin\theta\sin\varphi \;,
\end{multline}
where $\alpha$ is the constant of exchange interaction, $\theta$ and $\varphi$ are polar and azimuthal angles, respectively, of the spacer magnetic moment $\textbf{m}$, $\textbf{H}_m$ is the magnetostatic field in the system.

The value of the magnetostatic field can be easily derived from Maxwell's equation: $\text{div}\textbf{B}=\text{div}(\textbf{H}_m+4\pi \textbf{m})=0$. Since both the magnetization and therefore magnetostatic field depend only on one special variable, $z$, the magnetostatic field becomes: $\textbf{H}_m=-4\pi m_z\textbf{e}_z=-4\pi m \cos\theta\textbf{e}_z$, where $\textbf{e}_z$ is the unit vector along the $z$ axis.

Taking into account Eq.~(\ref{eq:4}), the Landau-Lifshitz equations in the angular form become
\begin{widetext}
\begin{subequations}
\begin{eqnarray} \label{subeq:5a}
\dfrac{\partial^2 \theta}{\partial \xi^2}=-\dfrac{d^2}{\Lambda^2}\left[\sin\theta\cos\theta+\sin\theta\dfrac{\partial\varphi}{\partial\tau}+\dfrac{H}{4\pi m}\cos\theta\cos\varphi+\dfrac{h}{4\pi m}\cos\theta\sin\varphi\right],
\end{eqnarray}
\begin{eqnarray}\label{subeq:5b}
\dfrac{\partial}{\partial\xi}\left(\sin^2 \theta\dfrac{\partial\varphi}{\partial\xi}\right)=
\dfrac{d^2}{\Lambda^2}
\left[\sin\theta\dfrac{\partial\varphi}{\partial\tau}+\dfrac{H}{4\pi m}\sin\theta\sin\varphi-\dfrac{h}{4\pi m}\sin\theta\cos\varphi \right].
\end{eqnarray}
\end{subequations}
\end{widetext}
The new dimensionless variables, normalized to the characteristic length and time in the problem, introduced in Eqs.~(\ref{subeq:5a}) and (\ref{subeq:5b}) are $\xi=z/d$ and $\tau=4\pi t \gamma m$, where $t$ is the time and $\gamma$ is the gyromagnetic ratio. $\Lambda=\sqrt{\alpha/4\pi}$ is the magnetic exchange length.\cite{18}

If the spacer thickness $d$ is much smaller than the magnetic exchange length $\Lambda$ ($d\ll\Lambda$), the right side in Eqs.~(\ref{subeq:5a}) and (\ref{subeq:5b}) becomes a small correction, which in the first approximation can be neglected.

As a result, only the exchange terms survive:  
\begin{subequations}\label{eq:6}
\begin{align}
\dfrac{\partial^2 \theta}{\partial \xi^2}&=0,\\
\dfrac{\partial}{\partial\xi}\left(\sin^2 \theta\dfrac{\partial\varphi}{\partial\xi}\right)&=0.
\end{align}
\end{subequations}

The solution, which satisfies the requirement of continuity of the polar and azimuthal components at the interfaces between the layers, has the form:
\begin{align}\label{eq:7}
\varphi(z)&=\varphi_2 +(\varphi_1 -\varphi_2)z/d,\notag\\ \varepsilon(z)&=\varepsilon_2 +(\varepsilon_1 -\varepsilon_2)z/d,\\
&0\leq z \leq d \notag.
\end{align}

The resulting magnetic energy of the spacer is
\begin{equation}\label{eq:8}
W =SJ\left[(\varphi_1-\varphi_2)^2+(\varepsilon_1-\varepsilon_2)^2\right]/2,
\end{equation}
where $J=\alpha m^2 / d=4\pi\Lambda^2m^2/d$.

To determine the resonance conditions for the layered system under consideration, we express the Lagrange function in terms of the angle:
\begin{multline}\label{eq:9}
L=\sum_{i=1}^2 \left(-S l_i\dfrac{M_i}{\gamma}\varepsilon_i\dfrac{\partial\varphi_i}{\partial t}-W_i\right)\\
+J\left[(\varphi_1-\varphi_2)^2+(\varepsilon_1-\varepsilon_2)^2 \right]/2.
\end{multline}

The variational equations following from Eg.~(\ref{eq:9}) are equivalent to the Landau-Lifshitz equations:
\begin{widetext}
\begin{equation}\label{eq:10}
\begin{pmatrix}
  iH_{\omega} & H+h_1 & 0 & -h_1\\
  4\pi M_1+H+h_1 & -iH_{\omega} & -h_1 & 0 \\
  0 & -h_2 & iH_{\omega} & H+H_{\text{b}}+h_2\\
  -h_2 & 0 & 4\pi M_2 +H+H_{\text{b}}+h_2 & -iH_{\omega}
\end{pmatrix}
\times
\begin{pmatrix}
\varepsilon_1 \\ \varphi_1 \\ \varepsilon_2 \\ \varphi_2
\end{pmatrix}
=
\begin{pmatrix}
h \\ 0 \\ h \\ 0
\end{pmatrix}
\end{equation}
\end{widetext}
where $h_i=J / M_i l_i=\alpha m^2 / d M_i l_i = 4 \pi \Lambda^2 m^2 / dM_i l_i $. Here, $H_{\omega}=\omega / \gamma$, $\omega = 2 \pi f$, $\gamma$ is the gyromagnetic ratio, $h_i$ is the characteristic field of exchange interaction between the layers.

The characteristic fields of the resonance modes of the collective spin dynamics in the system are found by equating the determinant of matrix (\ref{eq:10}) to zero. This results in two branches in the functional form of $H_{\omega}(H)$.

The first resonance branch, corresponding to the resonance field of F$_1$, has the form:
\begin{widetext}
\begin{multline}\label{eq:11}
H_{\omega}^2=(H_{\text{r1}}+h_1)(4\pi M_1+H_{\text{r1}}+h_1)
-h_1 h_2 \left[1-\dfrac{(2\pi M_1+H_{\text{r1}})(2\pi M_2 + H_{\text{r1}})}{2\pi H_{\text{r1}}(M_2 - M_1)}\right]\\
- h_1 h_2\dfrac{H_{\text{b}}}{H_{\text{r1}}}
\dfrac{4\pi\left[\pi(M_2+M_1)+ H_{\text{r1}}\right](M_2-M_1)-(2\pi M_1 + H_{\text{r1}})(2\pi M_2+H_{\text{r1}})}
{\left[ 2\pi(M_2-M_1)\right]^2},
\end{multline}
\end{widetext}
where $H_{\text{r1}}$ is the external field producing FMR in F$_1$ [see Fig.~\ref{fig2} (a)]. Only terms of order not higher than quadratic in $k_i$ were kept in Eq.~(\ref{eq:11}).

The value of $H_{\text{r1}}$ depends on whether the external magnetic field is parallel ($\uparrow\uparrow$) or antiparallel ($\uparrow\downarrow$) to the exchange bias field $H_{\text{b}}$. It is easy to show that the difference in the resonance fields, $\Delta H_{\text{r1}}=H_{\text{r1}}^{\uparrow\downarrow}-H_{\text{r1}}^{\uparrow\uparrow}$, has the form:
\begin{widetext}
\begin{eqnarray}\label{eq:12}
\Delta H_{\text{r1}}=H_{\text{r1}}^{\uparrow\downarrow}-H_{\text{r1}}^{\uparrow\uparrow}=
h_1 h_2\dfrac{H_{\text{b}}}{H_{r1}^0}
\dfrac{4\pi \left[\pi(M_2+M_1)+ H_{\text{r1}}^0 \right](M_2-M_1)-(2\pi M_1+H_{\text{r1}}^0)(2\pi M_2+H_{\text{r1}}^0)}
{(2\pi M_1+H_{\text{r1}}^0)\left[ 2\pi(M_2-M_1)\right]^2},
\end{eqnarray}
\end{widetext}
where $H_{\text{r1}}^0=(H_{\text{r1}}^{\uparrow\downarrow}+H_{\text{r1}}^{\uparrow\uparrow})/2=
\sqrt{(2\pi M_1)^2+H_{\omega}^{2}}-2\pi M_1-h_1$.

It follows from Eq.~(\ref{eq:12}) that $\Delta H_{\text{r1}}$ is proportional to a product of $h_1 h_2$. This means that $\Delta H_{\text{r1}}$ sharply changes in the vicinity of the Curie point of the spacer as a result of the sharp  increase in $m$ at the para-to-ferromagnetic transition (see Eq.~(\ref{eq:10}). Expectedly, $\Delta H_{\text{r1}}$ goes to zero as $T$ increases above the Curie point of the spacer. In this high-T limit, there is no coupling between F$_1$ and F$_2$, and Eq.~(\ref{eq:11}) describes the resonance field of the decoupled soft outer ferromagnet F$_1$. 

To find the resonance fields for F$_2$, we keep only terms of the order not higher than linear in $h_i$. The results for $H_{\text{r2}}^{\uparrow \downarrow}$ and $H_{\text{r2}}^{\uparrow\uparrow}$ are 
\begin{eqnarray}\label{eq:13}
H_{\text{r2}}^{\uparrow\uparrow} &=&
\sqrt{(2\pi M_2)^2 + H_{\omega}^2}-2\pi M_2-h_2-H_{\textbf{b}},
\notag\\
H_{\text{r2}}^{\uparrow\downarrow} &=&
\sqrt{(2\pi M_2)^2+H_{\omega}^2}-2\pi M_2-h_2+H_{\textbf{b}}.
\end{eqnarray}

Again, sharp changes in $H_{\text{r2}}^{\uparrow\uparrow}$ and $H_{\text{r2}}^{\uparrow\downarrow}$ are expected in the vicinity of the Curie point of the spacer. At high temperatures where $h_2 \rightarrow 0$, the difference between $H_{\text{r2}}^{\uparrow\downarrow}$ and $H_{\text{r2}}^{\uparrow\uparrow}$ naturally becomes $2H_{\text{b}}$.
 
It follows from Eq.~(\ref{eq:13}) that for sufficiently high values of $M_2$ and $h_2$, the $H_{\text{r2}}^{\uparrow\uparrow}$ branch can fall into negative fields, where it cannot be observed experimentally.

\section{Experiment}

\subsection{Samples and measurements}

The experiments were carried out on two sets of multilayered samples, in which either the spacer thickness, $d$, or its composition, $x$, were varied. The first set was Py(10~nm)/Ni$_{54}$Cu$_{46}$($d$)/Co$_{90}$Fe$_{10}$(5~nm)/ Mn$ _{80}$Ir$_{20}$(12~nm) (hereinafter -- F$_{1}$/Ni$_{54}$Cu$_{46}$($d$)/F$_2$/AF) with the spacer thicknesses $d=$ 3, 4.5, and 6 nm. The second set was Py(10 nm)/Ni$_x$Cu$_{100-x}$(6 nm)/Co$_{90}$Fe$_{10}$ (5 nm)/Mn$_{80}$Ir$_{20}$(12 nm) (hereinafter -- F$_1$/Ni$_x$Cu$_{100-x}$(6 nm)/F$_2$/AF), with $x=$ 54, 62 and 70 at.\%. The multilayers were deposited at room temperature on thermally oxidized silicon substrates using magnetron sputtering in an AJA Orion 8-target system. The exchange pinning between the ferromagnetic Co$_{90}$Fe$_{10}$ and antiferromagnetic Mn$_{80}$Ir$_{20}$ was set in during deposition of the multilayers using an in-plane magnetic field $H_{\text{dep}}\approx1$~kOe. Other fabrication details are similar to those described in Refs.~\onlinecite{7,8}.

In addition to the multilayers, single-layer Py (10 nm) and Co$_{90}$Fe$_{10}$ (5 nm) films were prepared under identical technological conditions. FMR measurements on the single-layer films were carried out to extract the magnetizations of Py and Co$_{90}$Fe$_{10}$ layers and use them for subsequent multilayer-FMR modelling and characterization [e.g., using Eqs.~(\ref{eq:12}) and (\ref{eq:13})].
  
The FMR measurements were performed using an X-band ELEXSYS E500 spectrometer equipped with an automatic goniometer. The operating frequency was $f=9.44$ GHz. FMR spectra for various in-plane dc-field angles were studied in the temperature range of 120 to 400 K.

\subsection{Results and discussion}
	
\subsubsection{Measured FMR spectra}

\begin{figure}[t,p]
\begin{center}
\includegraphics[width=7.5cm]{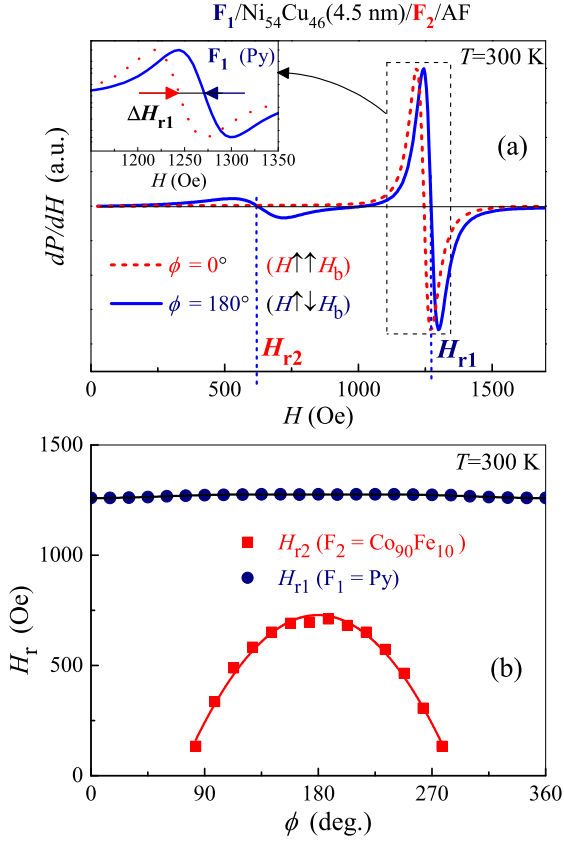}
\end{center}
\caption{(Color online)~(a) FMR spectra for F$_1$/Ni$_{54}$Cu$_{46}$(4.5 nm)f/F$_2$/AF for parallel (solid line) and antiparallel (red dashed line) orientations of the external magnetic field $H$ with respect to the exchange biasing field $H_{\text{b}}$. The upper inset shows an enlarged view of the signal from F$_1$. (b) The dependence of the resonance fields of F$_1$ and F$_2$ on the angle $\phi$ between $H$ and $H_{\text{b}}$ extracted from data sets such as those illustrated in (a).}
\label{fig2}
\end{figure}

Fig.~\ref{fig2} (a) shows two typical FMR spectra for a F$_1$/f/F$_2$/AF multilayer, for which the external magnetic field is parallel (solid line) or antiparallel (red dashed line) to the exchange bias field $H_{\text{b}}$ ($T=300$ K). The resonance signals from both F$_1$ and F$_2$ layers are clearly visible and are separated in field. As expected [see Eqs.~(\ref{eq:12}) and (\ref{eq:13})], the resonance conditions for both layers depend on the mutual orientation of $H$ and $H_{\text{b}}$ [Fig.~\ref{fig2}(b)]. Consistent with the predicted behavior of Eq.~(\ref{eq:13}), the $H_{\text{r2}}^{\uparrow\uparrow}$ branch for F$_2$ extrapolates into negative fields [see Fig.~\ref{fig2}(b)]. In the remainder of the paper we therefore discuss and in-depth analyze only the $H_{\text{r2}}^{\uparrow\downarrow}$ resonance branch as regards the dynamics of the pinned F$_2$ layer.

\begin{figure}[t,p]
\begin{center}
\includegraphics[width=7.5cm]{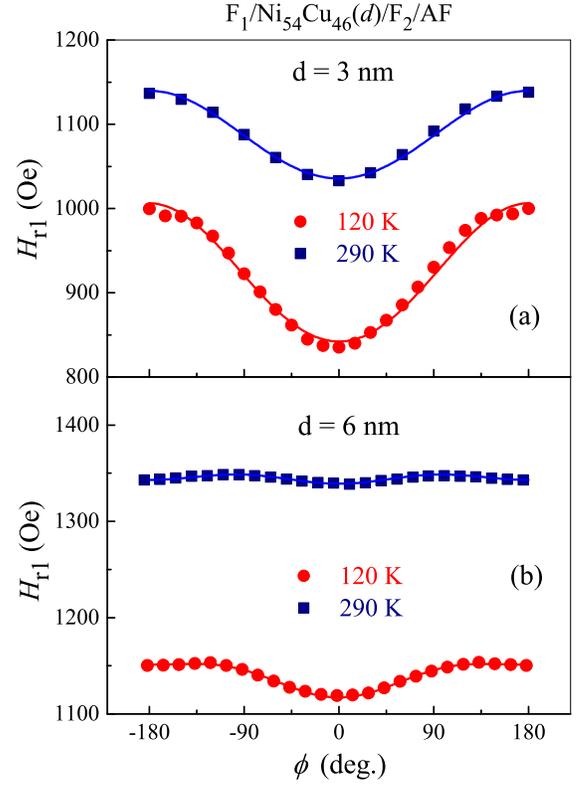}
\end{center}
\caption{(Color online)~Angular dependences of the resonance fields of the soft layer (F$_1$) in F$_1$/Ni$_{54}$Cu$
_{46}$($d$)/F$_2$/AF with the spacer thickness $d=$ 3 nm (a) and 6 nm (b).}  
\label{fig3}
\end{figure}

To understand the details of how the interlayer exchange coupling affects the spin dynamics of the free layer (F$_1$), the angular dependence of $H_{\text{r1}}$ was studied at various temperatures for various spacer thicknesses. The typical data shown in Fig.~\ref{fig3} indicate that the position of the resonance peak is angle-dependent and this angular asymmetry becomes stronger as the temperature is lowered. At the same time, the $H_{\text{r1}}$ vs $\phi$ dependence becomes more pronounced as the spacer thickness decreases. For all the cases shown in Figs.~\ref{fig2} (b) and \ref{fig3}, each $H_{\text{r1}}(\phi)$ data set is well fitted using a model characteristic of a thin film with unidirectional anisotropy (solid lines in Fig.~\ref{fig2}~(b) and Fig.~\ref{fig3}).

A more detailed analysis shows that there is an additional small contribution from uniaxial anisotropy. The extracted uniaxial anisotropy field ($\sim$4 Oe) is weakly dependent on temperature and spacer composition. This contribution potentially originates from some ordered configuration at NiCu/Py interface formed during the multilayer deposition in a magnetic field $H_{\text{dep}}$ (used for exchange pinning F$_2$).

\subsubsection{FMR modelling procedure}

\begin{figure}[t,p]
\begin{center}
\includegraphics[width=8cm]{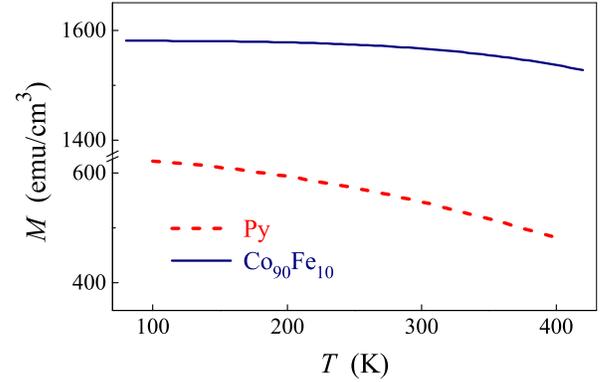}
\end{center}
\caption{(Color online)~Temperature dependences of the magnetizations of Py and Co$_{90}$Fe$_{10}$ layers obtained from the FMR data on the respective single-layered films.}
\label{fig4}
\end{figure}

Expressions (\ref{eq:12}) and (\ref{eq:13}) were used for calculating the temperature dependence of the resonance field asymmetry $\Delta H_{\text{r1}}=H_{\text{r1}}^{ \uparrow\downarrow}-H_{\text{r1}}^{ \uparrow\uparrow}$ for F$_1$ (Py), and of the resonance field  $H_{\text{r2}}^{\uparrow\downarrow}$ for F$_2$ (Co$_{90}$Fe$_{10}$). These expressions contain the values of the saturation magnetization $M_1$ and $M_2$ for the Py and Co$_{90}$Fe$_{10}$ layers, respectively, which are temperature dependent. The $M_1(T)$ and $M_2(T)$ dependences, used in the data analysis to follow, are shown in Fig.~\ref{fig4} and were obtained from the FMR data taken on single-layer Py (10 nm) and Co$_{90}$Fe$_{10}$ (5 nm) films prepared under the same conditions as the multilayers. The Kittel's formulas for isotropic thin films \cite{19,20} were used to calculate the $M_1(T)$ and $M_2(T)$ shown.

The key quantity determining the behavior of $\Delta H_{\text{r1}}$ and $H_{\text{r1}}^{\uparrow\downarrow}$, is the spacer magnetization $m$, averaged over the layer thickness [see Eq.~(\ref{eq:10})]. The spacer is ferromagnetic below the Curie point and nominally paramagnetic above it. It has previously been shown, however, that the proximity effect at the interface with a strong ferromagnet induces noticeable magnetization in a paramagnetic or weakly ferromagnetic metal and may give rise to an increase in its Curie point.\cite{8,21,22} The proximity length is an order of magnitude greater than the atomic spacing and for the case where the spacer thickness $d$ is of the order of a few nanometers, the induced magnetization penetrates through the spacer thickness.\cite{7,8} For this reason, to account for the proximity effect in our calculations, it was assumed that (i) there is an additional field $H_{\text{prox}}$ which acts on the spacer magnetization $m$ and (ii) for the spacer sandwiched between strong ferromagnets F$_1$ and F$_2$, the Curie point $T_{\text{C}}^{\text{eff}}$ differs from that in the bulk.

The $m(T,H_{\text{prox}})$ dependence due to the proximity effect was modelled using the mean-field approximation,\cite{14,23} where the magnetization is described by the Brillouin function:
\begin{align}\label{eq:14}
\dfrac{m}{m_0}=B_j(m)=\xi_1\cosh\left[\xi_1\left(\zeta_1\dfrac{m}{T}+\zeta_2\dfrac{H_{\text{prox}}}{T}\right)\right]\notag\\
-\xi_2\cosh\left[\xi_2\left(\zeta_1\dfrac{m}{T}+\zeta_2\dfrac{H_{\text{prox}}}{T}\right)\right];\notag \\
\xi_1=\dfrac{2j+1}{2j}; \xi_2=\dfrac{1}{2j};\zeta_1=\dfrac{3j}{j+1}\dfrac{T_{\textbf{C}}^{\text{eff}}}{m_0}; \zeta_2=\nu m_0;
\end{align}
where $j$ is the total angular momentum per Ni$_x$Cu$_{100-x}$ formula unit, $m_0$ is the saturation magnetization at $T=0$ K, $H_{\text{prox}}$ is the effective field reflecting the proximity effect at the interfaces with the strong ferromagnets F$_1$ and F$_2$, and $T_{\text{C}}^{\text{eff}}$ is the effective Curie temperature. Coefficient $\nu$ equals $\mu/(\rho N_{\text{A}})$, where $\mu$ and $\rho$ are the molar mass and density of Ni$_{x}$Cu$_{100-x}$, respectively, $N_{\text{A}}$ is the Avogadro constant, and $k_{\text{B}}$ is the Boltzmann constant.\cite{14,23}

The coefficient $\nu$ for our Ni$_x$Cu$_{100-x}$ alloy was estimated to be near $8.2 \times 10^{-8}$ cm$^{3}$ K erg$^{-1}$ for $x$ in the vicinity of 60 at.\%. The initial values of $j$ and $m_0$ were chosen based on the data calculated in Ref.~\onlinecite{7} for bulk Ni$_x$Cu$_{100-x}$, and the value of $j$ was kept fixed throughout the analysis. Since the magnetization and Curie temperature of the Ni$_x$Cu$_{100-x}$ spacer are expected to differ from those in the bulk, specifically due to the proximity effect, $m_0$ was chosen as one of the variable parameters in fitting the experimental data.

The proximity effect is expected to be most pronounced in the vicinity of $T_{\text{C}}^{\text{eff}}$. The \textit{ab-initio} calculations of this effect for F$_1$/Ni$_x$Cu$_{100-x}$($x,d$)/F$_2$/AF at $T \sim T_{\text{C}}^{\text{eff}}$ were detailed in Ref.~\onlinecite{8}. Based on a comparison of the values for the average magnetic moment $\langle m \rangle$ obtained in Ref.~\onlinecite{8} and the $m_{\text{calc}}(T)$ obtained using Eq.~\ref{eq:14}, it was found that $m_{\text{calc}}$ at $T \sim T_{\text{C}}^{\text{eff}}$ is approximately equal to $\langle m \rangle$ for $H_{\text{prox}} \approx$ 100 kOe. This value of $H_{\text{prox}}$ was kept fixed in all subsequent calculations.

Another important quantity affecting the spin dynamics in the system is the exchange bias field $H_{\text{b}}$. Based on the magnetometry measurements on F$_1$/Ni$_x$Cu$_{100-x}$($x,d$)/F$_2$/AF reported in Ref.~\onlinecite{7}, $H_{\text{b}}$ was obtained for a range of $x$ and $d$ values (for 300 K). These and the additional data reported in Ref.~\onlinecite{8} make it possible to conclude that for our samples with $x>52$, $H_{\text{b}}$ is only weakly temperature dependent. The calculation therefore assumed $H_{\text{b}}(T)=$ const. The specific fixed $j$ and $H_{\text{b}}$ values used in the calculations, among other parameters and variables, are presented in Table \ref{tab:1}. 
\begin{table*}
\caption{\label{tab:1}Magnetic parameters of F$_1$/Ni$_x$Cu$_{100-x}$($x,d$)/F$_2$/AF multilayers.}
\begin{ruledtabular}
\begin{tabular}{ccccccc}
$x$ (at.\% Ni) &$d$ (nm) &$j$ \footnotemark[1] &$H_{\text{b}}$ (Oe)\footnotemark[2] &$T_{\text{C}}^{\text{eff}}$ (K) & $m_0$ (emu/cm$^3$) &$\Lambda$ (nm)\\ 
\hline
0.54 &3.0 &0.19 &180 &450 &120 &11$\pm$2\\
0.54 &4.5 &0.19 &240 &320 &120 &11$\pm$2\\
0.54 &6.0 &0.19 &300 &250 &120 &11$\pm$2\\
0.62 &6.0 &0.21 &280 &350 &140 &12$\pm$2\\
0.70 &6.0 &0.23 &210 &550 &150 &13$\pm$2\\
\end{tabular}
\end{ruledtabular}
\footnotetext[1]{Values calculated from data of Ref.~\onlinecite{8} under assumption that Lande \textit{g}-factor equals 2.}
\footnotetext[2]{Values obtained from magnetic hysretesys loops at room temperature of Ref.~\onlinecite{7}.}
\end{table*}

Summarizing, the variable parameters used to fit the theoretical $\Delta H_{\text{r1}}(T)$ and $H_{\text{r2}}^{\uparrow\downarrow}(T)$ to the experimental data were the effective Curie temperature ($T_{\text{C}}^{\text{eff}}$) and saturation magnetization at $T=0$ ($m_0$) of the Ni$_x$Cu$_{100-x}$ spacer, as well as the characteristic magnetic exchange length ($\Lambda$). It will be shown below that for the case under study, the resulting values of $\Lambda$ are about two times greater than the spacer thickness $d$. This is within the limits of the approximation $d \ll \Lambda$ used in the analysis.

\subsubsection{FMR data analysis}

\begin{figure}[t,p]
\begin{center}
\includegraphics[width=7.5cm]{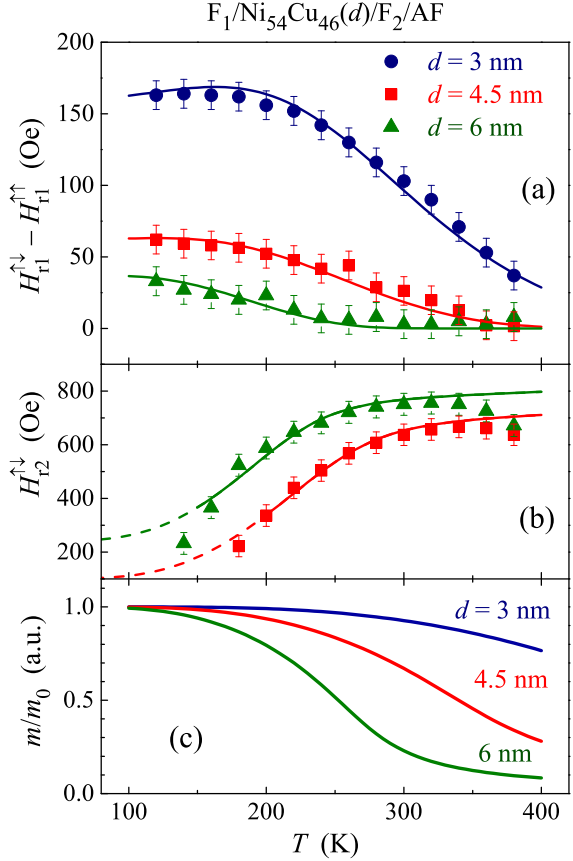}
\end{center}
\caption{(Color online)~Temperature dependences of $H_{\text{r1}}^{\uparrow\downarrow}-H_{\text{r1}}^{\uparrow\uparrow}$ (a) and $H_{\text{r2}}^{\uparrow\downarrow}$ (b) for F$ _1$/Ni$_{54}$Cu$ _{46}$($d$)/F$ _2$/AF samples for three spacer thicknesses (symbols). Bold solid lines show theoretical $H_{\text{r1}}^{ \uparrow\downarrow}-H_{\text{r1}}^{\uparrow\uparrow}$ and $H_{\text{r2}}^{\uparrow\downarrow}$ vs $T$ dependences fitted to the measured data using Eqs.~(\ref{eq:12}) and (\ref{eq:13}), respectively. (c) Temperature dependence of the normalized spacer magnetization $m / m_0$ obtained using the above fitting of the resonance fields.}
\label{fig5}
\end{figure}

Figs.~\ref{fig5} (a) and (b) show the temperature dependences of $H_{\text{r1}}^{\uparrow\downarrow}-H_{\text{r1}}^{\uparrow\uparrow}$ and $H_{\text{r2}}^{\uparrow\downarrow}$ for F$_1$/Ni$_{54}$Cu$_{46}$($d$)/F$_2$/AF samples obtained from the measured FMR spectra as well as the respective theoretical fits using the above analysis. A good agreement between the experiment and theory is obtained for realistic values of the fitting parameters.

The temperature dependence of the spacer magnetization $m/m_0$ obtained from fitting the resonance fields is shown in Fig.~\ref{fig5} (c) for different values of the spacer thickness. The proximity of the strongly ferromagnetic layers F$_1$ and F$_2$ has essentially no effect on the low-temperature magnetization of the spacer but is the dominant factor determining its effective Curie point $T_{\text{C}}^{ \text{eff}}$. The changes in $T_{\text{C}}^{\text{eff}}$ strongly depend on the spacer thickness: the smaller the $d$ and, therefore, the stronger the proximity effect of the interfaces, the higher the $T_{\text{C}}^{\text{eff}}$.

\begin{figure}[t,p]
\begin{center}
\includegraphics[width=7.5cm]{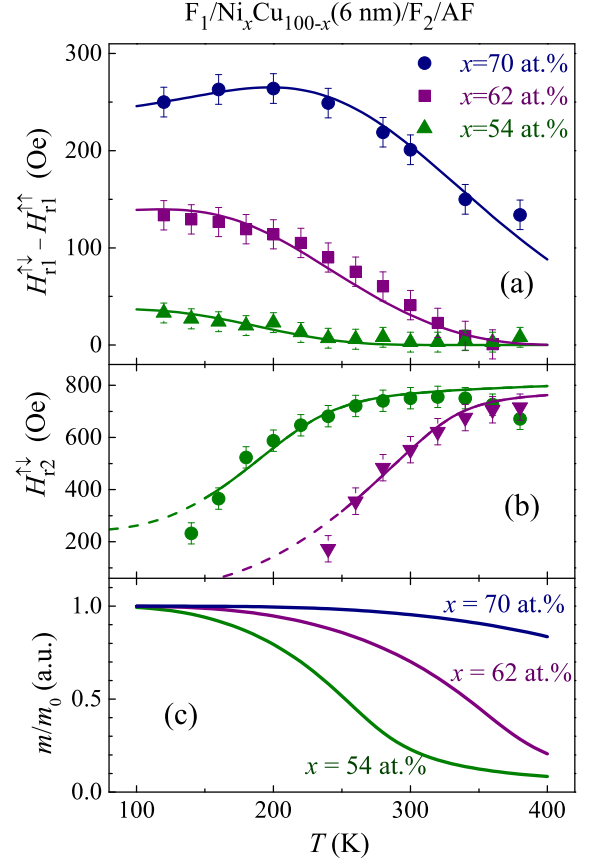}
\end{center}
\caption{(Color online)~Temperature dependences of $H_{\text{r1}}^{\uparrow\downarrow}-H_{\text{r1}}^{\uparrow\uparrow}$ (a) and $H_{\text{r2}}^{\uparrow\downarrow}$ (b) for F$ _1$/Ni$_x$Cu$_{100-x}$(6 nm)/F$_2$/AF samples for three Ni concentrations (symbols). Bold solid lines show theoretical $H_{\text{r1}}^{\uparrow\downarrow}-H_{\text{r1}}^{\uparrow\uparrow}$ and $H_{\text{r2}}^{\uparrow\downarrow}$ vs $T$ dependences fitted to the measured data using Eqs.~(\ref{eq:12}) and (\ref{eq:13}), respectively. (c) Temperature dependences of the normalized spacer magnetization $m/m_0$ obtained using the above fitting of the resonance fields.}
\label{fig6}
\end{figure}

Fig.~\ref{fig6} (a, b) shows the measured resonance fields $H_{\text{r1}}^{ \uparrow \downarrow} - H_{\text{r1}}^{\uparrow \uparrow}$ and $H_{\text{r2}}^{\uparrow \downarrow}$ as a function of temperature for F$_1$/Ni$_x$Cu$_{100-x}$(6 nm)/F$_2$/AF with different Ni-concentration in the spacer, fitted to theory using Eqs.~(\ref{eq:12}) and (\ref{eq:13}). The agreement is good, including the case of the highest Ni-concentration with non-monotonous temperature dependence of the resonance field asymmetry (70 at.\% Ni in Fig.~\ref{fig6} (a)). 

The dependence of the spacer magnetization on temperature extracted from fitting the data in Figs.~\ref{fig6} (a, b) is shown in Fig.~\ref{fig6} (c) for different Ni-concentration of the spacer. It is clear in this case that the proximity of the strongly ferromagnetic outer layers affects both the low-temperature magnetization $m_0$ and the effective Curie point $T_{\text{C}}^{\text{eff}}$ of the spacer -- the greater the $x$, the higher the $m_0$ and $T_{\text{C}}^{\text{eff}}$.

Having performed the data analysis, it is now informative to discuss the accuracy of the theoretical assumption made in Section 2 in describing the spin dynamics in an F$_1$/f/F$_2$/AF spin-thermionic system. One key assumption was that the magnetization vectors of the F$_1$ and F$_2$ layers are parallel to the external field $H$ and perform only weak oscillation under the influence of the alternating field component $h$. This assumption is strictly correct for the case where $H$ is parallel to $H_{\text{b}}$, but the opposite antiparallel case ($H \uparrow \downarrow H_{\text{b}}$) must be considered with care. 

The values of $H_{\text{b}}$ for F$_1$/Ni$_x$Cu$_{100-x}$($x,d$)/F$_2$/AF multilayers are listed in Table~\ref{tab:1}. As follows from the $M-H$ data presented in Refs.~\onlinecite{7,8}, the magnetization of the F$_1$/f/F$_2$/AF multilayer fully saturates when the applied reversing field $H^{\uparrow \downarrow}$ exceeds (1.2 $\div$ 1.5) times $H_{\text{b}}$. Thus, the above assumption should be valid when both resonance fields, $H_{\text{r1}}^{\uparrow \downarrow}$ and $H_{\text{r2}}^{\uparrow \downarrow}$, are higher than (1.2 $\div$ 1.5)$H_{\text{b}}$. For the samples under study, $H_{\text{r1}}^{\uparrow \downarrow}$ is greater than 900 Oe, so the first condition, $H_{\text{r1}}^{\uparrow \downarrow}>(1.2 \div 1.5)H_{\text{b}}$, is well satisfied. The data in Figs.~\ref{fig5} and \ref{fig6} indicate that the second condition, $H_{\text{r2}}^{\uparrow\downarrow}>(1.2 \div 1.5)H_{\text{b}}$, is also well satisfied for temperatures above $\sim$ 180 K.

The developed approach, combining theory and experiment, makes it possible to extract and analyze the thickness and composition dependence of the characteristic parameters of the spacer, which are summarized in Table~\ref{tab:1}. 

\begin{figure}[t,p]
\begin{center}
\includegraphics[width=7.5cm]{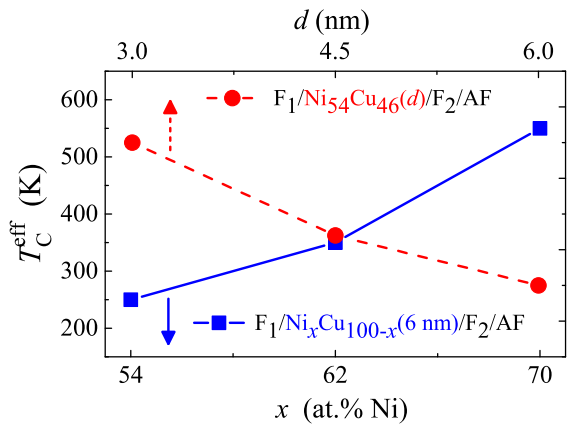}
\end{center}
\caption{(Color online)~Dependence of the effective Curie point $T_{\text{C}}^{\text{eff}}$ of the spacer Ni$_x$Cu$_{100-x}$ on its thickness $d$ (at fixed composition $x=54$ at.\% Ni) and Ni content $x$ (at fixed thickness $d=6$ nm).}
\label{fig7}
\end{figure}

Fig.~\ref{fig7} presents the model-extracted dependence of the effective Curie point $T_{\text{C}}^{\text{eff}}$ of the spacer on its thickness $d$ and Ni content $x$. We conclude that small variations in both $d$ and $x$ result in significant variations in $T_{\text{C}}^{\text{eff}}$. We also note that the values of the Curie temperature of Ni$_x$Cu$_{100-x}$ sandwiched between strong ferromagnets are much greater than the corresponding values in the bulk form of this diluted magnetic alloy -- ($T_{\text{C}}^{\text{bulk}} \approx$ 120 K and 300 K for $x=$ 54 and 70 at.\% Ni, respectively)\cite{9}.    

\section{Conclusion}

Ferromagnetic resonance properties of F$_1$/f/F$_2$/AF multilayers, where spacer f has a low Curie point compared to strongly ferromagnetic F$_1$ and F$_2$, have been analyzed theoretically and investigated experimentally. The spacer-mediated exchange coupling is shown to strongly affect the resonance fields of both F$_1$ and F$_2$ layers. It is found that the key magnetic parameters of the spacer which govern the magnetic resonance in the system are the magnetic exchange length ($\Lambda$), the effective saturation magnetization at $T=0$ $(m_0)$, and the effective Curie temperature ($T_{\text{C}}^{\text{eff}}$) of the spacer.

By theoretically fitting the measured FMR data, the values of $\Lambda$, $m_0$, and $T_{\text{C}}^{\text{eff}}$ are obtained for the technologically significant ranges in Ni-concentration ($x=54 \div 70$ at.\% Ni) and thickness ($d=3 \div 6$ nm) of the spacer. The values thus obtained are entirely different from the corresponding quantities in the bulk. The developed approach to spin dynamics in the system enables such detailed quantitative characterization that is otherwise is difficult or impossible obtain in terms of direct measurements due to the built-in strong proximity effect.

The inferred magnetism in the key element of the structure -- the spacer, acting as an interlayer exchange-spring -- shows a great sensitivity and thereby high tunability of its properties versus the degree of magnetic dilution, geometry, and temperature. These results should be useful for designing high-speed nanodevices based on spin-thermionic control.

\section*{Acknowledgments}
Support from the Swedish Stiftelse Olle Engkvist Buggm\"{a}stare, Swedish Research Council (VR grant 2014-4548), the Science and Technology Center in Ukraine (project P646), and the National Academy of Sciences of Ukraine (projects 0115U003536 and 0115U00974) are gratefully acknowledged.

\end{document}